\documentclass[%
  superscriptaddress,
  reprint,
 amsmath,amssymb,
 aps,
]{revtex4-2}

\usepackage{orcidlink}
\usepackage{graphicx}% Include figure files
\usepackage{dcolumn}% Align table columns on decimal point
\usepackage{bm}% bold math
\usepackage{upgreek} % Adds in support for greek letters in roman typeset
\usepackage{mathrsfs}
\usepackage{xspace} 

\def\SND {\mbox{SND@LHC}\xspace}

\begin{document}

%\preprint{APS/123-QED}

\title{Observation of collider neutrinos without final state muons with the \SND experiment}% Force line breaks with \\

\author{D.~Abbaneo~\orcidlink{0000-0001-9416-1742}}
\affiliation{European Organization for Nuclear Research (CERN), Geneva, Switzerland}
  
\author{S.~Ahmad~\orcidlink{0000-0001-8236-6134}}
\affiliation{Currently at: Pakistan Institute of Nuclear Science and Technology (PINSTECH), Nilore, 45650, Islamabad Pakistan}

\author{R.~Albanese~\orcidlink{0000-0003-4586-8068}}
\affiliation{Sezione INFN di Napoli, Napoli, Italy}
\affiliation{Universit\`{a} di Napoli ``Federico II'', Napoli, Italy}

\author{A.~Alexandrov~\orcidlink{0000-0002-1813-1485}}
\affiliation{Sezione INFN di Napoli, Napoli, Italy}

\author{F.~Alicante~\orcidlink{0009-0003-3240-830X}}
\affiliation{Sezione INFN di Napoli, Napoli, Italy}
\affiliation{Universit\`{a} di Napoli ``Federico II'', Napoli, Italy}

\author{K.~Androsov~\orcidlink{0000-0003-2694-6542}}
\affiliation{Institute of Physics, \'{E}cole Polytechnique F\'{e}d\'{e}rale de Lausanne (EPFL), Lausanne, Switzerland}

\author{A.~Anokhina~\orcidlink{0000-0002-4654-4535}}
\affiliation{Affiliated with an institute covered by a cooperation agreement with CERN}

\author{T.~Asada~\orcidlink{0000-0002-2482-8289}}
\affiliation{Sezione INFN di Napoli, Napoli, Italy}
\affiliation{Universit\`{a} di Napoli ``Federico II'', Napoli, Italy}

\author{C.~Asawatangtrakuldee~\orcidlink{0000-0003-2234-7219}}
\affiliation{Chulalongkorn University, Bangkok, 10330, Thailand}

\author{M.A.~Ayala Torres~\orcidlink{0000-0002-4296-9464}}
\affiliation{Center for Theoretical and Experimental Particle Physics, Facultad de Ciencias Exactas, Universidad Andr\`es Bello, Fernandez Concha 700, Santiago, Chile}

% ...

\collaboration{The \SND Collaboration}%\noaffiliation

\date{\today}% It is always \today, today,

\begin{abstract}
We report the observation of neutrino interactions without final state muons at the LHC, with a significance of 6.4\,$\sigma$. A data set of proton-proton collisions at 
$\sqrt{s}= 13.6$\,TeV collected by \SND in 2022 and 2023 is used, corresponding to an integrated luminosity of 
68.6\,fb$^{-1}$.
Neutrino interactions without a reconstructed muon are selected, resulting in an event sample consisting mainly of neutral-current and electron neutrino charged-current interactions in the detector.
After selection cuts, 9 neutrino interaction candidate events are observed  with an estimated background of 0.32 events.
\end{abstract}

\maketitle

%\tableofcontents

%\section{Introduction}
\label{sec:intro}
{\it Introduction -} 
The \SND experiment consists of a detector 
placed in the TI18 tunnel at a distance of 480\,m,
in the forward region, from the ATLAS detector located in the Interaction Point 1 (IP1) of the CERN Large Hadron Collider (LHC).
The detector is designed to detect and measure interactions of neutrinos produced 
in decays of particles produced in proton-proton 
collisions at IP1. The neutrinos produced in the 
forward direction stem from high-energy hadrons 
and have energies of a few hundred GeV to several
TeV\cite{Beni:2019gxv,Beni:2020yfy,Kling:2021gos}, leading to deep inelastic scattering (DIS) in the experiment.

Data taking at the LHC started in 2022, and in Ref.~\cite{SNDLHC:2023pun} a direct
observation of muon neutrino charged current (CC) interactions  was reported based on a data sample 
of 36.8\,fb$^{-1}$. A similar observation was 
reported in Ref.~\cite{FASER:2023zcr} and a measurement of electron and muon neutrino cross sections using an emulsion detector was reported in Ref.~\cite{FASER:2024hoe}.

In this paper, we search for neutrino interactions
with no muons in the final state, 
referred to as $\nu0\mu$ events. This measurement
is sensitive to CC interactions of
either $\nu_{e}$ or $\nu_{\tau}$ 
flavour, and neutral current (NC) neutrino interactions.
The observation of a clear signal above background is a sign of other than $\nu_{\mu}$ CC neutrino interactions at the LHC.
Hence this analysis is the first step towards flavour classification  with \SND. The strategy is to select muon-less shower-like events, 
but the analysis is not optimised to distinguish between $\nu_{e}$, $\nu_\tau$ and NC events. 
The data sample used, collected in 2022 and 2023, corresponds to a total recorded integrated luminosity of 
68.6\,fb$^{-1}$. 

%\section{The \SND detector}
\label{sec:detector}

{\it The \SND detector -} 
 \SND is a hybrid detector consisting of emulsion and electronic detectors\cite{SNDLHC:2022ihg}. It is optimized for the detection of interactions of all neutrino flavors. 
 The electronic detectors provide the time stamp of the neutrino interaction and pre-select the interaction region while the neutrino-interaction vertex is reconstructed using tracks in the tracking detectors and the emulsion target. The Veto system is used to tag muons and other charged particles entering the detector from the IP1 direction.
 
The Veto system consists of two parallel planes of scintillating bars. Each plane is made of seven $1 \times 6 \times 42$\,cm$^3$ stacked horizontal bars of plastic scintillator.
The target section contains five walls. Each wall consists of four units (`bricks’) of Emulsion Cloud Chambers (ECC) with tungsten as target material, and is followed by a scintillating fiber (SciFi) station for tracking.
Each SciFi station consists of one horizontal and one vertical $39 \times 39$\,cm$^2$ plane. Each plane comprises six staggered layers of 250\,$\upmu\textrm{m}$ diameter polystyrene-based scintillating fibers. The single particle spatial resolution in one plane is roughly
100\,$\upmu\textrm{m}$ and the time resolution for a particle crossing both $x$ and $y$ planes is about 250 ps.

The muon system consists of two parts: the first five stations (Upstream, US), and the last three stations (Downstream, DS).  
The eight stations are interleaved with 20 cm-thick iron blocks. Each US station consists of 10 stacked horizontal scintillator bars of $82.5 \times 6 \times 1$\,cm$^3$, resulting in a coarse $y$ view and
together they 
act mainly as a hadronic calorimeter. A DS station consists of two layers of thinner bars measuring $82.5 \times 1 \times 1$\,cm$^3$, arranged in alternating $x$ and $y$ planes, leading to  a spatial resolution in each coordinate of less than 1\,cm for 
muon reconstruction and a time resolution of about 150\,ps. Hits in the DS detector and the SciFi tracker are used to identify events with muons.
All signals exceeding pre-set thresholds are read out by the front-end electronics and clustered in time to form events. An efficient software noise filter is applied online to the events, resulting in negligible detector dead time and negligible loss in signal efficiency. Events satisfying certain topological criteria, such as the presence of hits in several detector planes, are written to disk. In the absence of beam, the noise filtering logic reduces the event rate by five orders of magnitude to around 4 Hz. At the highest instantaneous luminosity in 2022 and 2023 ($2.5 \times 10^{34}$\,cm$^{-1}$\,s$^{-1}$), this setup generated a rate of around 5.4 kHz.

This analysis does not use the information from the emulsion detector but is based on the hits and 
track reconstruction in the scintillating fiber 
detectors and on the absence of an identified muon
in the muon system.
 
%\section{Data and simulation}
\label{sec:data}

{\it Data and simulation -} 
The data used for the analysis described in this letter was taken between July 2022 and October 2023, comprising the first two years of the LHC Run\,3. An integrated luminosity of 68.6\,fb$^{-1}$ was recorded by the experiment's electronic detectors during this period. A data collection efficiency of 97.2\% was achieved relative to the luminosity delivered at IP1 as reported by the ATLAS experiment\cite{ATLAS:2022hro,ATL-DAPR-PUB-2024-001}. We highlight the exceptional performance of the detector in 2023, with 31.8\,fb$^{-1}$ collected and an uptime of 99.7\%. There were six exchanges of the emulsion-instrumented target during the data taking period. The mass of tungsten installed in the target during this period was 792\,kg.

For two periods comprising about half of the collected luminosity, the event builder was not aligned with the timing of bunch crossings at the interaction point. This results in occasional splitting of events and as a consequence the efficiency of the veto detector is worse in these periods. The impact of the lower veto efficiency on this analysis is negligible, as downstream components of the detector are also used to reject the dominant penetrating background due to muons.

Data taken in a hadron test beam in 2023 in the CERN North Area are used to validate the analysis methods. The test-beam detector consisted of four SciFi stations smaller in size than those of \SND, an exact replica of the US hadron calorimeter, and a single DS station. Iron blocks were inserted between the SciFi stations with a thickness equivalent in interaction lengths to the target walls in the \SND detector. The test-beam detector was exposed to hadron beams of mostly pions with momenta ranging from 100 to 300\,GeV/c.

Proton-proton collision events are simulated with the \textsc{DPMJET-III} generator\cite{Roesler_2001,DPMJET}, which is interfaced with a \textsc{Fluka}\cite{Fluka2,Fluka3} model of the LHC\cite{Boccone:2014hxd}. Particles are propagated to a virtual plane around 60\,m upstream of the \SND detector and recorded for further processing. Neutrino interactions in the detector are simulated with the \textsc{Genie} generator\cite{GENIE}. Muon DIS interactions in the tunnel material are simulated with \textsc{Pythia6}\cite{Pythia6}. The particles resulting from these interactions are propagated through a \textsc{Geant4}\cite{Geant4} model of the tunnel and the detector geometry implemented in the \textsc{FairRoot} framework\cite{al_turany_2024_11210174}. The same setup is used for the test-beam simulation.

The sample of simulated neutrino interactions comprises 3$\times10^{5}$ events, corresponding to an integrated luminosity of 40\,ab$^{-1}$. The computational expense of simulating muon DIS interactions and propagating the large number of produced particles through the tunnel and detector models, compounded with the very low probability of these interactions being selected into the analysis samples, results in a significant computing challenge. Out of 3$\times 10^{8}$ simulated muon DIS interactions, corresponding to an integrated luminosity of 20\,fb$^{-1}$, none survive the event selection described in the following section. This sample is sufficient to set a stringent upper limit on the number of events expected from this source of background, but it does not allow for probing the detector's response to this type of event. For this purpose, we generate additional samples of 2.6$\times 10^7$ neutrons and neutral kaons impinging on the upstream face of the target, with energy spectra extracted from the muon DIS simulation. We note that according to the muon DIS simulation, no neutral hadrons above 100\,GeV are expected to interact in the target in the absence of Veto system hits due to charged particles accompanying the neutral hadrons.
 
%\section{Event selection and data analysis methods}
\label{sec:analysis}
\begin{figure}[t]
    \centering
    \includegraphics[width=0.5\textwidth]{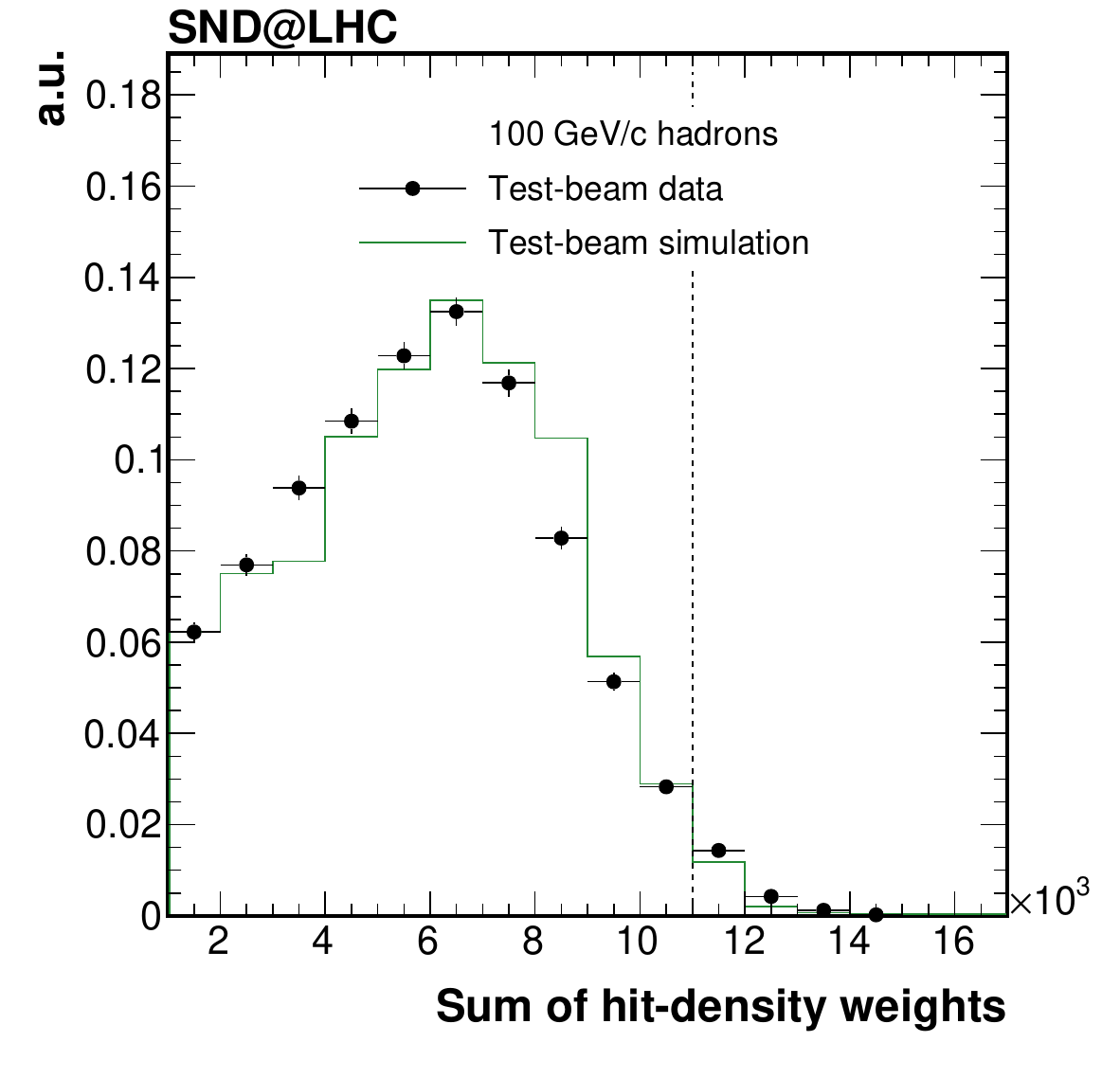}
    \caption{Distribution of the sum of SciFi hit-density weights for 100\,GeV/c test-beam hadrons.}
    \label{fig:TB100GeV}
    \end{figure}

{\it Event selection and data analysis methods -} 
The search for the $\nu0\mu$ signal proceeds in two stages. The first stage consists of a set of selection criteria to isolate events that are consistent with neutral particles interacting in the tungsten target that do not produce reconstructible muons. In the second stage, a discriminating variable is used to define regions enriched in either signal or neutral hadron background. The background-rich region is used to constrain the background expectation, and the signal-rich region is optimised by maximising the expected significance for excluding the background-only hypothesis.

The analysed events are required to have occurred during LHC stable beam conditions and in coincidence with a proton bunch pair colliding at IP1. In order to avoid complications arising from Veto detector dead time, events are further required to occur at least 625\,ns after the previous recorded event. After applying this set of quality criteria the data set comprises 1.1$\times 10^{10}$ events, with a signal efficiency greater than 99.75\%.

Events consistent with a neutral particle interacting in the target are then selected by requiring that the average position in the 
transverse $xy$ plane of the SciFi hits lies in a tight fiducial area (roughly $25 \times 25$\,cm$^2$). As an additional criterion to remove events due to particles entering the top and bottom parts of the detector, it is required that there are no hits in the uppermost and lowermost two bars of the first two layers of the hadron calorimeter. In order to remove events resulting from particles entering the detector from the IP1 direction we require no hits in the veto system. These requirements discard the most common events in the data, associated with charged particles entering the detector from the sides and from the upstream end. The fiducial cuts result in a signal acceptance of 20\% and reduce the data set to 7.9$\times 10^{6}$ events.
\begin{figure*}[t]
    \includegraphics[width=0.94\textwidth]{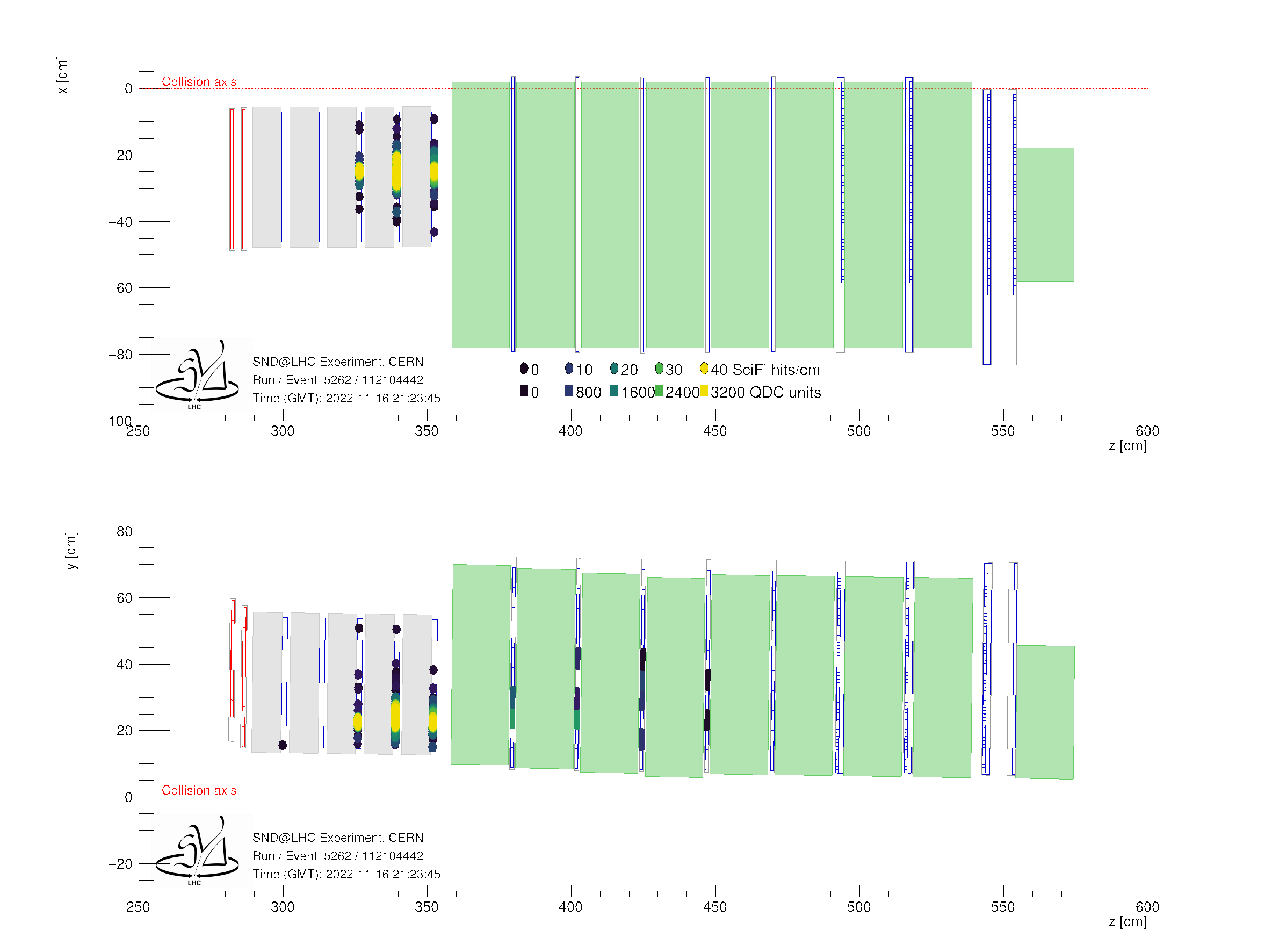}
    \caption{Representative example of a signal-like event. The top panel shows a top-down view of the detector and the bottom panel shows the view from the side. The coloured circles represent the local density of hits in the SciFi detector, corresponding to the number of hits within 1 cm of each hit. The coloured rectangles represent the amplitude, in arbitrary units, of hits in the US hadron calorimeter. Lighter shades correspond to higher values.}
    \label{fig:eventDisplay}
\end{figure*}

To ensure adequate sampling of the electromagnetic and hadronic shower components, events are further required to span at least two SciFi stations and to produce hits in the two most upstream layers of the hadron calorimeter. A minimum of 35 SciFi hits is required, as well as large activity in the hadron calorimeter, with an energy deposition roughly equivalent to at least 10 GeV. These cuts further reduce the data to 2.3$\times 10^{4}$ events, while keeping 58\% of the remaining signal.

In order to exclude reconstructible muons, events with hits in the two most downstream planes of the DS detector are also discarded. This ensures that fewer than three DS planes have hits, and no muon tracks can be reconstructed. The background due to $\nu_\mu$ CC interactions is reduced by a factor of 14. The residual $\nu_\mu$ CC background is dominated by events where the muon exits the detector through the sides.

The second step of the event selection makes use of a variable developed to identify events with a high density of hits in one SciFi station, as would be expected for events with large electromagnetic or hadron showers. A weight $w_i$ is assigned to each SciFi hit $i$ which is calculated by counting the number of other hits in the same detector plane within 1 cm:
\begin{displaymath}
    w_i = \sum_{j \ne i}^{N} H(x_j-x_i+\textrm{1~cm}) \times \left[ 1 - H(x_j-x_i-\textrm{1~cm}) \right]
\end{displaymath}
\noindent where $N$ is the number of hits in a SciFi plane, $x_i$ are the positions of the hits in cm, and $H(x)$ is the Heavyside step function. The hit weights can vary from zero, when no neighbouring hits are present, to 80 when all adjacent channels are activated.

The weights of all hits are added up separately for each SciFi station. To discard events with sparse hit patterns, it is required that the sum of hit-density weights is larger than 20 in at least two stations, and larger than 2$\times 10^3$ in at least one station. This reduces the dataset to 25 events, with 18.7 neutrino interactions expected.

The highest sum of hit-density weights among the five SciFi stations is then used as a discriminating variable to separate high-energy neutrino interactions from low-energy neutral hadrons. This variable was validated using hadron test-beam data. The distribution of the sum of hit-density weights for 100\,GeV/c hadrons interacting in the test-beam detector target is shown in Figure~\ref{fig:TB100GeV}, where it can be seen that most neutral hadrons of this energy produce SciFi hit-density weight sums smaller than 11$\times 10^3$.

A control region comprising events with summed SciFi hit-density weights smaller than 5$\times 10^3$ is used to estimate the background due to neutral hadrons resulting from muon DIS interactions in the tunnel walls. Ten events are observed in the control region, with 3.1 neutrino interactions expected. This is consistent with the 90\% CL upper limit of 8.2 events obtained from the muon DIS simulation. We use the high-statistics neutral-hadron simulation to extrapolate this constraint to higher SciFi hit-density sum regions.

The signal region is defined by estimating the sensitivity to observe $\nu0\mu$ events for different thresholds of the hit-density weight sum, ranging from 5$\times 10^3$ to 20$\times 10^3$, and choosing the threshold that maximises this metric. The sensitivity is quantified in terms of the exclusion of the null hypothesis, defined by setting the $\nu0\mu$ signal strength, $\alpha$, to zero. The one-sided profile likelihood ratio test $\lambda(\alpha)$ is used as the test statistic. The sensitivity is calculated by comparing the test statistic evaluated at the nominal expectation from the simulation, $\lambda_{\textrm{MC}}(\alpha=0)$, with the sampling distribution of $\lambda(\alpha=0)$. The likelihood, which includes a Gaussian factor to account for the background uncertainty, is
\begin{displaymath}
\mathscr{L} = \mathrm{Poisson}(n\,|\, \alpha \cdot s + \beta)\times \mathrm{Gauss}(\beta\,|\,b,\sigma_\beta)
\end{displaymath}
where $n$ is the number of events in the signal region, $s$ is the expected number of signal events and $\beta$ is the number of background events given by the Gaussian model, having a mean value $b$ and a standard deviation $\sigma_\beta$. We use the RooStats\cite{rootstats} implementation of this method.

A systematic uncertainty of 100\% is assigned to the neutral hadron background, to account for the extrapolation of the control region constraint to the signal region. The systematic uncertainty on the background due to $\nu_\mu$ CC interactions is 18.2\%. This uncertainty is given by the statistical uncertainty on the number of events observed in the dedicated analysis\cite{SNDLHC:2023pun, SNDMoriond24}, added in quadrature with the DS tracking systematic uncertainty estimated with muon tracks tagged in the SciFi detector\cite{SNDLHC:2023mib}. An uncertainty of 100\% is assigned to the background due to $\nu_\tau$ CC interactions with a muon in the final state.

\begin{table*}[t]%The best place to locate the table environment is directly after its first reference in text
    \begin{ruledtabular}
    \begin{tabular}{lccc}
    \textrm{}&
    \textrm{Data}&
    \textrm{$\nu0\mu$ Simulation}&
    \textrm{$\nu1\mu$ Simulation}\\ \colrule
    %\colrule
    All events                        & $1.11\,\times\,10^{10}$       & 307.7  & 495.8 \\ 
    Fiducial volume                   & $7.91\,\times\,10^{6}$        & 63.6   & 111.7  \\ 
    Hits in 2 SciFi and 2 US stations & $5.95\,\times\,10^{5}$        & 38.0   & 83.4  \\
    Large SciFi activity              & $6.34\,\times\,10^{4}$        & 37.1   & 74.5  \\
    Large hadron calorimeter activity               & $2.28\,\times\,10^{4}$        & 35.8   & 68.3  \\
    No hits in last 2 DS stations     & $1.47\,\times\,10^{4}$        & 27.7   & 4.9   \\
    Sparse-shower removal             & 25                            & 16.7   & 2.0   \\
    Signal region                     & 9                             & 7.2    & 0.30  
    \end{tabular}
    \end{ruledtabular}
    \caption{\label{tab:cutflow} Number of events passing the selection cuts in the data, and in neutrino signal ($\nu0\mu$) and background ($\nu1\mu$) simulation.}
\end{table*}

The threshold of summed hit-density weights that maximises the sensitivity to observe the $\nu0\mu$ signal is 11$\times 10^3$. In this signal region, the total systematic uncertainty on the number of expected background events is 17.9\%. A representative example of a signal-like event is shown in Figure~\ref{fig:eventDisplay} and a summary of the number of events passing each selection cut is given in Table~\ref{tab:cutflow}.

%\section{Results}
\label{sec:results}

{\it Results -} 
The expected significance of the $\nu0\mu$ observation is 5.5\,$\sigma$, with 7.2 signal events expected over a background of 0.30 $\nu_\mu$ CC events, 1.5$\times 10^{-2}$ neutral hadron events resulting from muon DIS interactions in the tunnel walls, and 1.7$\times 10^{-3}$ $\nu_\tau$ CC interactions with a muon in the final state. The signal expectation is composed of 4.9 $\nu_e$ CC events, 2.2 NC events, and 0.1 $\nu_\tau$ CC events with no muons in the final state.

\begin{figure}[t]
\centering
\includegraphics[width=0.5\textwidth]{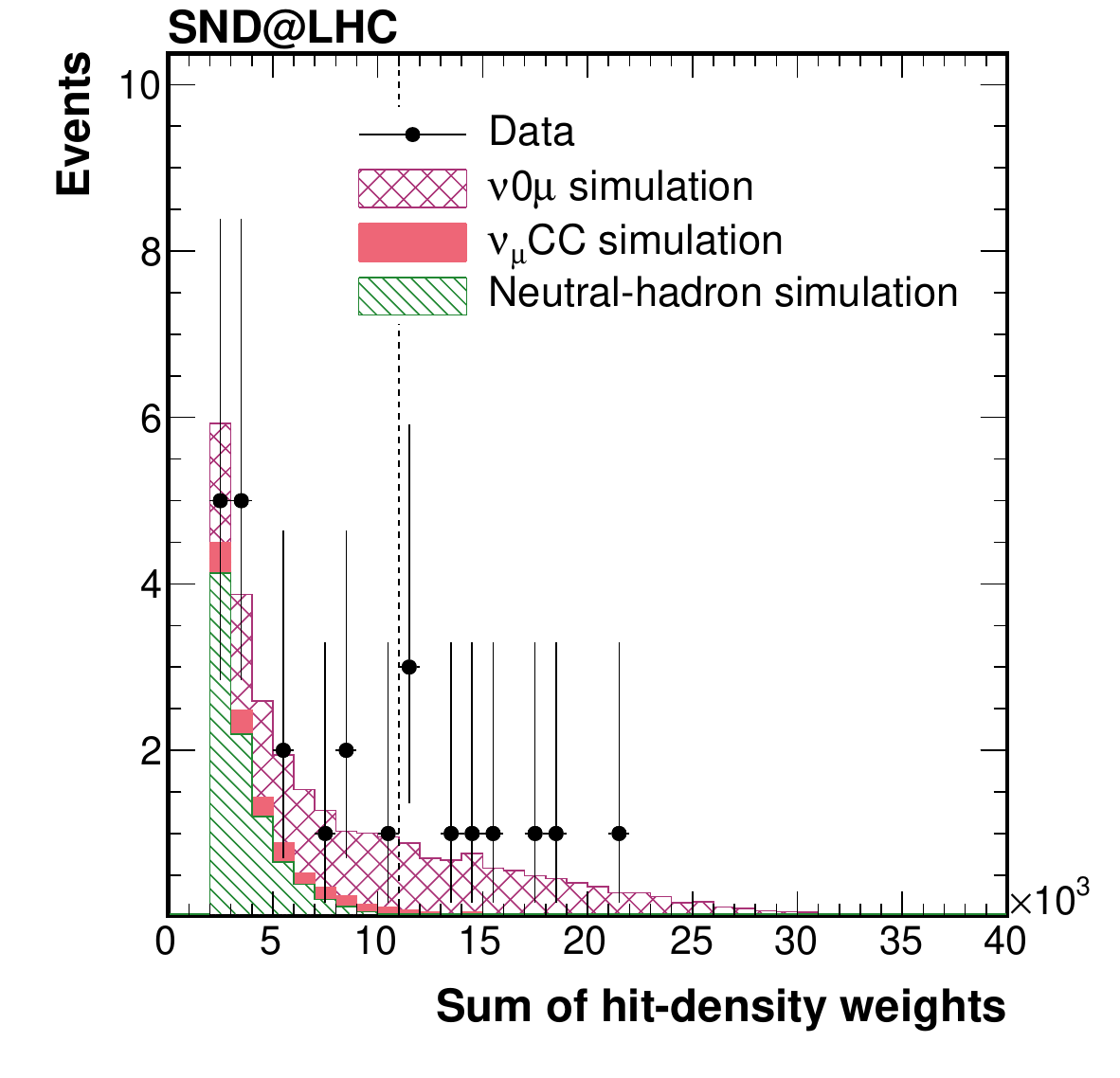}
\caption{Distribution of the sum of SciFi hit-density weights for events selected into the analysis sample. The events from the data are shown alongside the expected signal and background.}
\label{fig:nudata}
\end{figure}

Nine events are observed in the signal region, resulting in an observation significance of 6.4\,$\sigma$. An example of a signal-like event is shown in Figure~\ref{fig:eventDisplay}. The distribution of the sum of SciFi hit-density weights for events observed in the data is shown in Figure~\ref{fig:nudata}, along with signal and background expectations.

We also consider the significance to observe $\nu_e$ CC interactions with this analysis. The NC component of the $\nu0\mu$ signal originates primarily from the flux of $\nu_\mu$. Therefore, we can constrain the NC component of the $\nu0\mu$ signal using the measurement of $\nu_\mu$ CC interactions in Ref.~\cite{SNDLHC:2023pun}. With this constraint, we extract evidence for the observation of $\nu_e$ CC events at the level of 3.7\,$\sigma$, with an expected significance of 2.2\,$\sigma$. The uncertainties on the $\nu_\mu$ CC and NC components are conservatively taken to be fully correlated.

%\section{Conclusions}
\label{sec:conclusions}

{\it Conclusions -} 
We report the observation of neutrino interactions without final state muons at the LHC, with a significance of 6.4\,$\sigma$ (5.5\,$\sigma$ expected), and evidence for $\nu_e$ CC interactions with a significance of 3.7\,$\sigma$ (2.2\,$\sigma$ expected). The selected event sample consists mostly of NC and $\nu_e$ CC interactions. Taken with the previously reported observation of $\nu_\mu$ CC interactions, this result is the first demonstration of flavour classification by the \SND experiment. We expect that minor modifications of the method used for this analysis, when applied to the large amount of data collected by the experiment during 2024, will be sensitive to the direct observation of $\nu_e$ CC events using the electronic detectors of \SND. A parallel effort is underway to select these events in the emulsion detector data.

{\it Acknowledgments -} We express our gratitude to our colleagues in the CERN accelerator departments for the excellent performance of the LHC. We thank the technical and administrative staffs at CERN and other institutes  for their contributions to the success of this effort. We acknowledge the support for the construction and operation of the detector provided by the following funding agencies:  CERN; the Bulgarian Ministry of Education and Science within the National
Roadmap for Research Infrastructures 2020–2027 (object CERN); the German Research Foundation (DFG, Deutsche Forschungsgemeinschaft); the Italian National Institute for Nuclear Physics (INFN); JSPS, MEXT, the Global COE program of Nagoya University, the Promotion
and Mutual Aid Corporation for Private Schools of Japan for Japan;
the National Research Foundation of Korea with grant numbers 2021R1A2C2011003, 2020R1A2C1099546, 2021R1F1A1061717, and 
2022R1A2C100505; Fundação para a Ciência e a Tecnologia, FCT (Portugal), CERN/FIS-INS/0028/2021; the Swiss National Science Foundation (SNSF); TENMAK for Turkey (Grant No. 2022TENMAK(CERN) A5.H3.F2-1). This work has been partially supported by Spoke 1 ``Future HPC \& BigData" of ICSC - Centro Nazionale di Ricerca in High-Performance-Computing, Big Data and Quantum Computing, funded by European Union - NextGenerationEU. M.~Climescu and R.~Wanke are funded by the Deutsche Forschungsgemeinschaft (DFG, German Research Foundation), project 496466340. We acknowledge the funding of individuals by Fundação para a Ciência e a Tecnologia, FCT (Portugal) with grant numbers CEECIND/01334/2018, CEECINST/00032/2021 and PRT/BD/153351/2021. We thank Luis Lopes, Jakob Paul Schmidt and Maik Daniels for their help during the construction. 
%\nocite{*}

\bibliography{sn-bibliography}% Produces the bibliography via BibTeX.

\end{document}